# Assimilating Observed Surface Pressure into ML Weather Prediction Models

**L. C. Slivinski[1], J. S. Whitaker[1], S. Frolov[1], T. A. Smith[1], N. Agarwal[1,2]**

[1]NOAA/OAR Physical Sciences Laboratory, Boulder, CO, USA.
[2]CU Boulder Cooperative Institute for Research in Environmental Sciences, Boulder, CO, USA.
Corresponding author: Laura Slivinski (laura.slivinski@noaa.gov)

## Abstract

There has been a recent surge in development of accurate machine learning (ML) weather prediction models, but evaluation of these models has mainly been focused on medium-range forecasts, not their performance in cycling data assimilation (DA) systems. Cycling DA provides a statistically optimal estimate of model initial conditions, given observations and previous model forecasts. Here, real surface pressure observations are assimilated into several popular ML models using an ensemble Kalman filter, where accurate ensemble covariance estimation is essential to constrain unobserved state variables from sparse observations. In this cycling DA system, deterministic ML models accumulate small-scale noise until they diverge. Mitigating this noise with a spectral filter can stabilize the system, but with larger errors than traditional models. Perturbation experiments illustrate that these models do not accurately represent short-term error growth, leading to poor estimation of cross-variable covariances.

## 1 Introduction

In recent years, many pure and hybrid machine learning (ML)-based weather forecast models have been developed with impressive results. For example, the pure emulators FourCastNet (Pathak et. al. 2022), PanguWeather (Bi et. al. 2023), and GraphCast (Lam et. al. 2023) have all shown performance comparable to state-of-the-art operational weather prediction models. In addition, Kochkov et. al. (2024) introduced NeuralGCM, a hybrid model that uses a differentiable dynamical core with physics parameterizations learned from a neural network that can make ensemble weather forecasts comparable to state of the art operational ensembles and can be run stably for decades. Bonev et. al. (2023) introduced Spherical Fourier

Neural Operators (SFNOs) which improve upon the original FourCastNet architecture's 10-day forecast accuracy, and enable stable, year-long simulations. The European Centre for Medium Range Weather Forecasts (ECMWF) has developed a weather forecasting emulator that is currently being run in parallel to their operational forecast system (Lang et. al. (2024)).

If these models are to be trusted to produce operational forecasts, their performance needs to be quantified better than using simple root mean squared error (RMSE) forecast metrics. For example, Hakim and Masanam (2024) showed that PanguWeather produces realistic signal propagation from several canonical perturbation experiments, suggesting that it does encode realistic physics. On the other hand, Selz and Craig (2023) found that PanguWeather does not accurately capture the rapid error growth from small-scale perturbations, the so-called "butterfly effect". Tian et. al. (2024) investigated the tangent linear and adjoint models of GraphCast and NeuralGCM and found unphysical behavior, particularly in vertical cross-sections of these models. Related, Bonavita (2024) has shown that several popular ML models do not reproduce the energy spectra of physical models, especially at longer forecast lead times, and argues that this implies that these models do not completely represent dynamically balanced relationships. These models therefore need to be better understood before their use beyond medium-range forecasts.

Evaluation of these models has generally been focused on forecast skill, often at 1-10 day lead times. However, operational weather prediction systems rely on data assimilation (DA) systems to provide initial conditions for the forecast model. For example, the Artificial Intelligence/integrated Forecast System (AIFS) that is run at ECMWF is still initialized using their traditional cycling DA system. If these ML models are ever to successfully replace physical models in a fully cycled DA system, they need to be able to handle assimilation of observations. In addition, DA can provide important information about the model that may not be gleaned from forecasts alone (e.g. Dee & Da Silva, 1998).

Several studies have attempted to use ML models within a cycling DA system. Xiao et. al. (2023) successfully developed a 4DVar system with their FengWu model, but only assimilated synthetic gridded observations from ERA5. Similarly, Adrian et. al. (2024) assimilated partial ERA5 data into FourCastNet. Other studies have attempted to learn the DA step itself, either via so-called "end-to-end" forecasting, which takes only observations as input and outputs a full field weather prediction (e.g. Vaughan et. al., 2024), or by combining an ML forecast model with an ML DA step as in FuXi Weather (Sun et. al. 2024). FuXi Weather assimilated real satellite and GNSS-RO observations into a background field generated with the FuXi model. They showed improvements in forecast skill of some variables relative to ECMWF's HRES, but only beyond day 3-7. Recently, the FuXi model was integrated into an ensemble 4DVar system, FuXi-En4DVar (Li et. al., 2024). The authors argue that this system produces physically meaningful analysis increments, though only synthetic observations from ERA5 were assimilated.

Here, we present results from a stringent test of several popular ML models in a cycling DA framework. In particular, we only assimilate real surface pressure observations from around the globe using a pure ensemble Kalman filter (EnKF). Previous work on historical reanalysis (Compo et. al. 2011, Slivinski et. al. 2019) has shown that assimilating only surface pressure observations into a modern atmospheric weather model can produce analysis fields with skill comparable to a 3-4-day modern weather forecast. Compo et. al. (2006) also showed that accurately capturing the dynamic covariances from an ensemble DA system is crucial for spreading the information from sparse observations to unobserved variables.

Assimilating real surface pressure observations into these models is an efficient way to expose possible deficiencies in these models, not just within a cycling framework but also within forecast frameworks. In particular, results from these experiments may point to issues when using these models for forecasting, and help inform development of future models intended specifically for operational forecasting.

## 2 Methods

### 2.1 Cycling DA experiment setup

Several ML models were tested in a cycling DA framework in which only surface pressure observations are assimilated with an EnKF. Traditional serial DA consists of a forecast step and an analysis step. Here, the goal is to test the effect of replacing the physical model in the forecast step with a pure or hybrid ML model. Specifically, an 80-member ensemble forecast is extended to 6h using either a physical model or an ML model. Surface pressure observations are assimilated into this forecast ensemble to provide an updated analysis ensemble. The analysis ensemble is used to initialize the forecast for the next assimilation cycle, and so on. Note that the DA analysis step does not use any ML techniques in these experiments. The DA update was performed with an EnKF, which does not require an adjoint from any of the models, nor is a static background error covariance needed. Rather, the background error covariance is estimated directly from the forecast ensemble.

Two versions of a hybrid model and three pure emulators were compared to a traditional Numerical Weather Prediction (NWP) model. NeuralGCM, a hybrid model, has a deterministic and a stochastic version, both of which have 32 vertical levels. "NeuralGCM-det" has a horizontal resolution of 0.7 degrees, but is generally not stable for forecasts longer than about a week (Hoyer, pers. comm.). "NeuralGCM-stoch" is a stochastic forecast model that can run stably for decades, and has a horizontal resolution of 1.4deg. Unlike NeuralGCM-det, which uses a traditional Mean Squared Error (MSE) loss function, NeuralGCM-stoch includes a Continuous Ranked Probability Score (CRPS) term in its training, as well as Gaussian Random Fields in the forward step of the emulator for stochasticity. GraphCast, PanguWx, and SFNO are all recently-released pure emulators that predict either surface pressure or mean sea level pressure (MSLP). These emulators all have a horizontal resolution of 0.25deg and 13 vertical levels, and were optimized to produce accurate medium-range forecasts. These five models were compared to the medium-range weather application of NOAA's Unified Forecast System (UFS), which uses the finite-volume cubed-sphere dynamical core Global Forecast System (FV3GFS; Lin 2004, Putman & Lin 2007) at a resolution of C96 (approx. 1 deg) with 127 hybrid sigma-pressure vertical levels. The UFS employs stochastic physics and, at a higher resolution, is used for operational ensemble prediction at NOAA.

Experiments were initialized from 29 Aug 2021, 18UTC from the 80-member operational UFS ensemble, interpolated to the relevant horizontal resolution, and cycled for two months (or until the model diverged). Surface pressure observations were assimilated every 6 hours. NeuralGCM provides forecasts at 3h lead time, so for those comparisons, three 3-hourly backgrounds were used (beginning, middle, and end of the 6h window). For comparisons to GraphCast, PanguWx, and SFNO, only one background in the middle of the window was used, since these models provide forecasts at a minimum length of 6h.

At each analysis step, surface pressure observations from the operational Global Data Assimilation System (GDAS) BUFR files were assimilated using an EnKF to update wind, temperature, and humidity on global rectangular grids and on vertical model levels. Specifically, the serial ensemble square root filter (EnSRF; Whitaker & Hamill 2002) was configured as in the 20th Century Reanalysis version 3 (Slivinski et. al. 2019), except that no observation bias correction or nonlinear quality control was used here. Adaptive horizontal and vertical localization and relaxation to prior spread (RTPS) inflation (Whitaker & Hamill 2012) were used. The forward observation operator was the MAPS pressure reduction from model terrain to station height (Benjamin & Miller 1990). For GraphCast and PanguWx, model terrain was set to 0 since the updated variable is mean sea level pressure (MSLP) instead of surface pressure. SFNO includes both surface pressure and MSLP as state variables, but only surface pressure was used in the forward operator of the DA since tests (not shown) using MSLP in the forward operator degraded performance.

2.2 Perturbation experiment setup

In addition to the cycling experiments, a perturbation experiment was performed to evaluate the short-term error growth in these models. Two of the pure emulators, GraphCast and SFNO, were compared to the UFS. Two 80-member ensembles were produced as perturbations of each other: one ensemble had no observations assimilated, and the second ensemble had one surface pressure observation assimilated located in the North Pacific (57N, 182E). Six-hour forecasts were computed from each model, and resulting differences were analyzed. The perturbation from the assimilation of a single observation was used to probe how the models respond to an adjustment relevant to DA.

# 3 Results

## 3.1 Cycling DA experiments

Within four weeks of cycling with real surface pressure observations, all pure ML models and NeuralGCM-det became unstable (Figure 1). NeuralGCM-stoch was the only model that did not diverge, though its errors in 500hPa geopotential height (Z500) are about 20% worse than the UFS. Z500 errors from SFNO are comparable to those from NeuralGCM-stoch until it diverged. GraphCast and PanguWx show rapidly growing errors before they diverged. NeuralGCM-det initially gave errors smaller than the UFS (possibly due to its higher horizontal resolution), but the model diverged within two weeks of cycling. Note that the initial Z500 error growth in all experiments is due to the relatively accurate initial conditions; spinup is expected since assimilating surface pressure observations cannot provide operational-quality analyses.

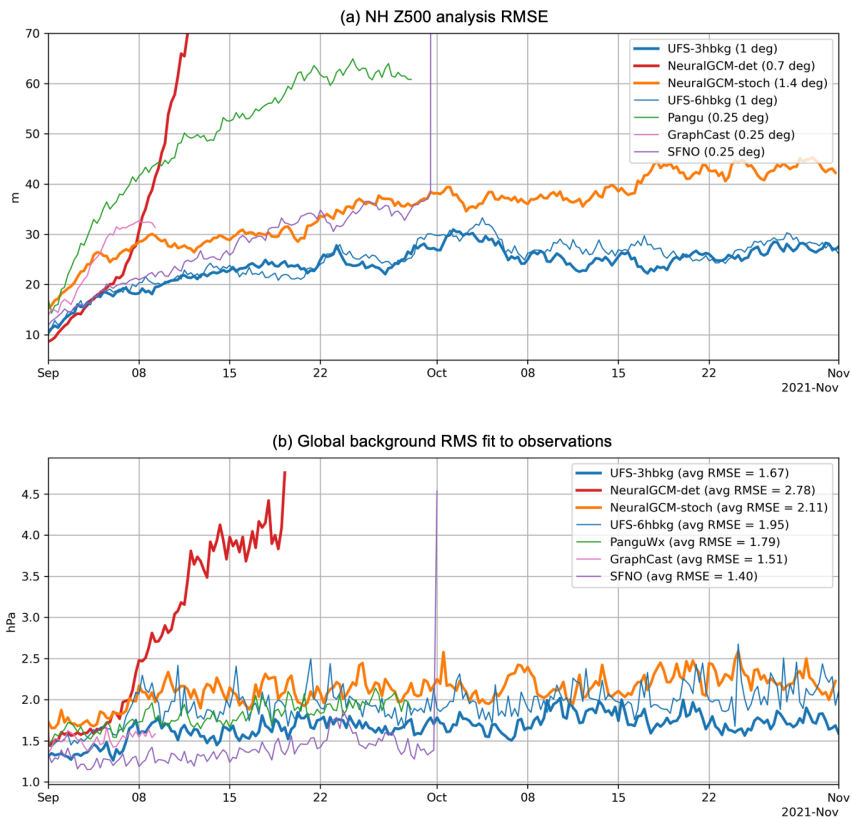

**Figure 1.** Time series of errors from each cycling experiment. Thick curves use three backgrounds per analysis cycle (3-hourly backgrounds), and thin curves use one background per analysis cycle (6-hourly backgrounds). **(a)** Root-mean square analysis error of 500hPa geopotential height (Z500), relative to ERA5, in the Northern Hemisphere (20N-90N). Curves end when the models become unstable. **(b)** Root-mean square difference between observations and background fields interpolated to observation time and location. Average RMSE for the available time is given in the legend.

The growing instabilities are illustrated in the MSLP/surface pressure increments from each model for several different cycles (Figure S1). The pure emulators show growing high frequency signals at near-grid scale, suggesting that the covariances do not correctly spread out information from observations. This small-scale noise grew until the models diverged. The increments from the hybrid NeuralGCM-det show the classic numerical instability signal of high-frequency wave-like structures initiating in the tropics. Increments from the UFS grew in magnitude as a result of spinup from assimilating only a sparse observation network.

The growth of high-frequency noise in these models is further illustrated in the 700hPa kinetic energy spectra of the first-guess ensemble means (Figure 2). NeuralGCM-det, GraphCast, and SFNO had growing energy at small scales, while the spectra of UFS and NeuralGCM-stoch were relatively stable in time. Interestingly, PanguWx did not show the same growth in small-scale energy, although its increments became noisy (Fig. S1).

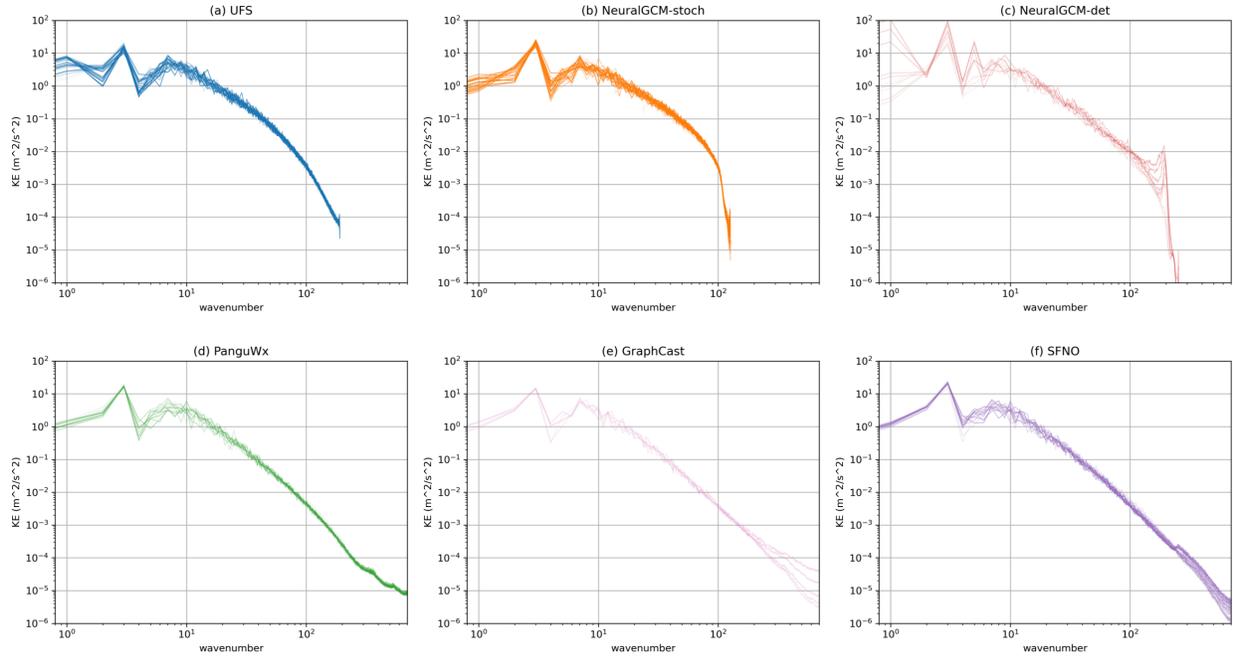

**Figure 2.** Instantaneous global kinetic energy spectra at 700hPa of the background ensemble mean for each experiment. Each curve illustrates spectra from a different DA cycle; lighter colors denote earlier dates in the experiment, and darker colors denote later dates.

      To test whether damping this small-scale energy would improve the stability of these models, a Gaussian spectral filter with *e*-folding scale 80 was applied to the increments in GraphCast. This stabilized the cycling and prevented the growth of small-scale energy (Figure 3). However, the Z500 errors are nearly three times those of the UFS and continue to grow after 60 days, even as the background fits to surface pressure observations are comparable to those from the UFS (similar to the unfiltered system). This suggests that the instability of the pure emulators is not the root cause of their poor performance, but rather that they do not accurately estimate background ensemble covariances.

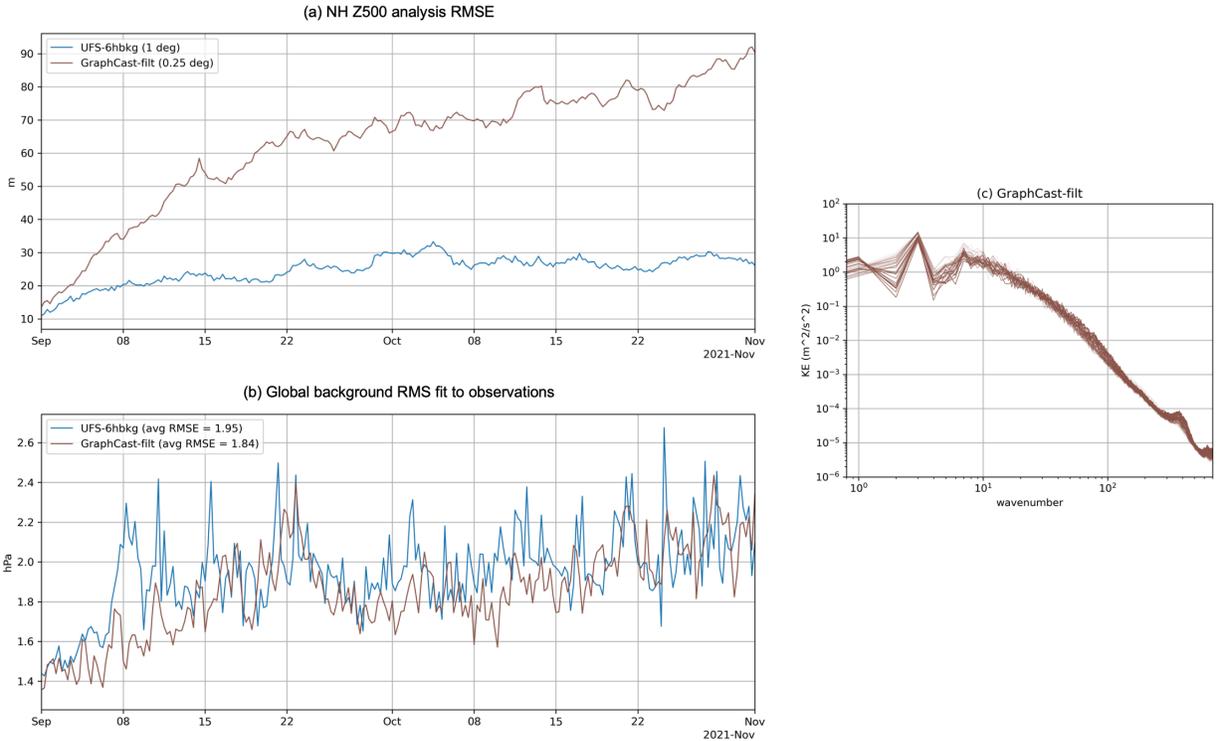

**Figure 3. (a-b)** As in Figure 1, but for the filtered GraphCast experiment. **(c)** As in Figure 2, but for the filtered GraphCast experiment.

This hypothesis is supported by maps of ensemble correlation between surface pressure/MSLP and Z500 (Fig S2). The correlations from GraphCast, SFNO, and PanguWx degraded during cycling, relative to the UFS. In particular, even when the increments were filtered, the correlations from GraphCast quickly became too weak. Correlations from neuralGCM-stoch did not degrade in time and are generally close to 1, similar to the UFS. While this only illustrates local correlations across two variables, the results support the hypothesis that the full cross-covariances from the pure emulators are inaccurate.

3.2 Perturbation experiments

The pure emulators likely have inaccurate estimates of cross-covariances because they do not accurately represent growth of small-scale perturbations. Figure 4 illustrates this by showing ensemble mean differences in Z500 from the perturbation experiment described in Section 2.2. As expected, there is a gravity wave in the UFS emanating from the perturbation location. SFNO has a weak signal of a gravity wave, while GraphCast does not exhibit a gravity wave at all. Surprisingly, other than the gravity wave, the horizontal structure of the ensemble mean differences in the region of the perturbation are similar in all three models. However, the vertical extent and magnitude of the difference is much stronger in GraphCast and SFNO than in UFS. This suggests that GraphCast and SFNO cannot accurately capture the propagation of ensemble perturbations over short time periods necessary for the estimation of covariances that are responsible for the performance of ensemble DA. This results in inaccurate and noisy increments that are fed back into the emulators. Since these models seem to have no way to restore balance, they diverge, even as the fields fit the observations well (Fig. 1).

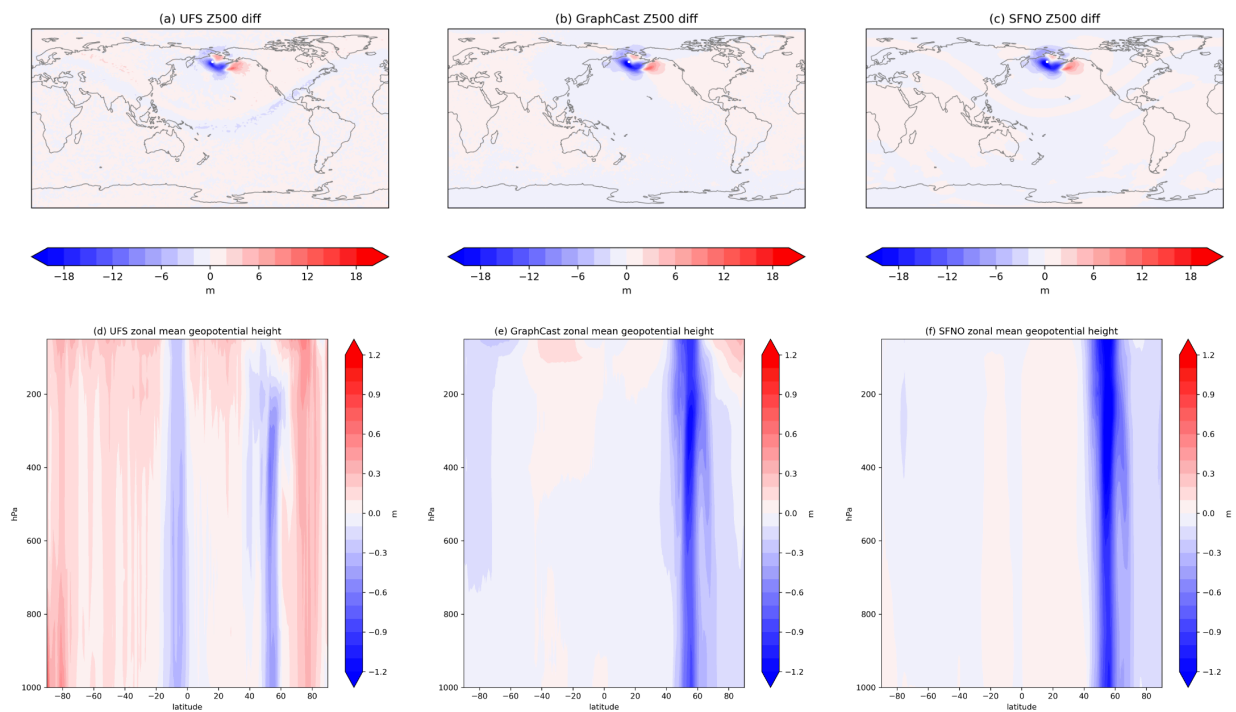

**Figure 4**. Ensemble mean differences from perturbation experiments. Top row: Maps of Z500 ensemble mean difference. Bottom row: Vertical profiles of ensemble mean zonal mean geopotential height difference.

## 4 Conclusions

Recent advances in deep learning weather forecasting show promise in potentially changing the paradigm of operational weather forecasting. For instance, most operational forecast centers run some version of an ensemble-based prediction system, which could benefit from much larger ensembles than are currently possible with physics-based models. However, these models need to accurately represent the covariance structures in order to be useful for this purpose. At the same time, confronting models (ML or physics-based) with observations via DA can efficiently reveal issues in the models. To test the limit of some top-performing ML models, we replaced the physics-based NWP model with ML models in a cycling surface-pressure-only DA system. This experiment is relatively easy to set up and only requires that the models predict surface pressure or MSLP. However, it is also a strict test of the models' ability to estimate covariances necessary for spreading information across variables.

We tested several state-of-the-art ML models that are currently publicly available and that estimate either surface pressure or MSLP. NeuralGCM-stoch is the only model tested that was able to cycle stably for over a month without modifications, but it produced Z500 errors 30-40% worse than the UFS. Further work would be needed to determine whether and how to modify NeuralGCM-stoch to produce errors comparable to those from the UFS (by, e.g., increasing horizontal resolution). NeuralGCM-det and the pure emulators all became unstable. Spectral filtering of the noisy surface pressure increments stabilized the cycling, but produced errors that are more than twice those from the UFS and are comparable to the skill of model climatology.

Recall that the results shown here used surface pressure in the forward operator for SFNO, not MSLP as in GraphCast and PanguWx. An experiment cycling SFNO in which MSLP was used produced worse errors, comparable to those from PanguWx, and diverged more quickly (not shown). This suggests that these models could benefit from including the variables to be updated by DA in their training, although other issues remain since both tests with SFNO eventually diverged.

Single-observation experiments showed that GraphCast and SFNO cannot evolve perturbations representative of small-scale errors correctly. Furthermore, GraphCast lacks representation of gravity waves. These results suggest that training these models to produce the best forecast RMSEs does not necessarily yield accurate estimation of spatial and inter-variable correlations, and therefore the pure ML models studied here cannot accurately represent the background error covariances necessary for cycling DA. Instead, repeatedly adding DA increments to these models increases the energy at small scales, which never gets transferred to larger scales (e.g. Selz & Craig, 2023), resulting in the eventual model divergence. In addition, local correlations between variables seem to get weaker and weaker until there is effectively no information transferred from the observations to the rest of the state. Moreover, the vertical correlation structure in the data driven models is very different from what is shown by the UFS. Namely, the vertical correlation in the ML models is too strong and extends too high, in agreement with results from Tian et. al. (2024).

At present, most ML models have been optimized for short- to medium-range weather forecasts. Out of the box, the deterministic ML models studied here cannot be used in a cycling DA system. In particular, these models cannot be naively used to generate very large ensembles to improve DA since they do not simulate accurate correlations. GenCast (Price et. al. 2024) offers an attractive alternative for producing ensembles, but at present it can only produce 12h forecasts and therefore cannot be evaluated in this framework. We suggest that ML models will need to be produced specifically for the application of DA, likely using different training criteria than are currently used for training pure forecast models. For example, FuXi Weather, the system that assimilates real observations into the FuXi model, was finetuned on an earlier version of FuXi used for DA (Sun et. al. 2024). We hypothesize that models trained for DA may need to be optimized for short-term forecasts (6 hours or less) while simultaneously exhibiting long term stability. Future model development efforts should also focus on how to capture cross-variable covariances, mitigate growth of small-scale energy, and preserve physical balance when perturbations are propagated. Finally, we note that these models will need to include additional variables than they do at present in order to be applicable in an operational NWP context in which satellite radiance would be assimilated.

**Acknowledgments**

The authors would like to acknowledge support from the NOAA Physical Sciences Laboratory.

**Open Research**

These experiments used software available from Putman & Lin (2007); Bi et. al. (2023); Lam et. al. (2023); Kochkov et. al. (2024); and Bonev et. al. (2023). Initial conditions are available at https://registry.opendata.aws/noaa-gfs-bdp-pds/ and observations are available at https://psl.noaa.gov/data/nnja_obs/.

**Supporting Information**

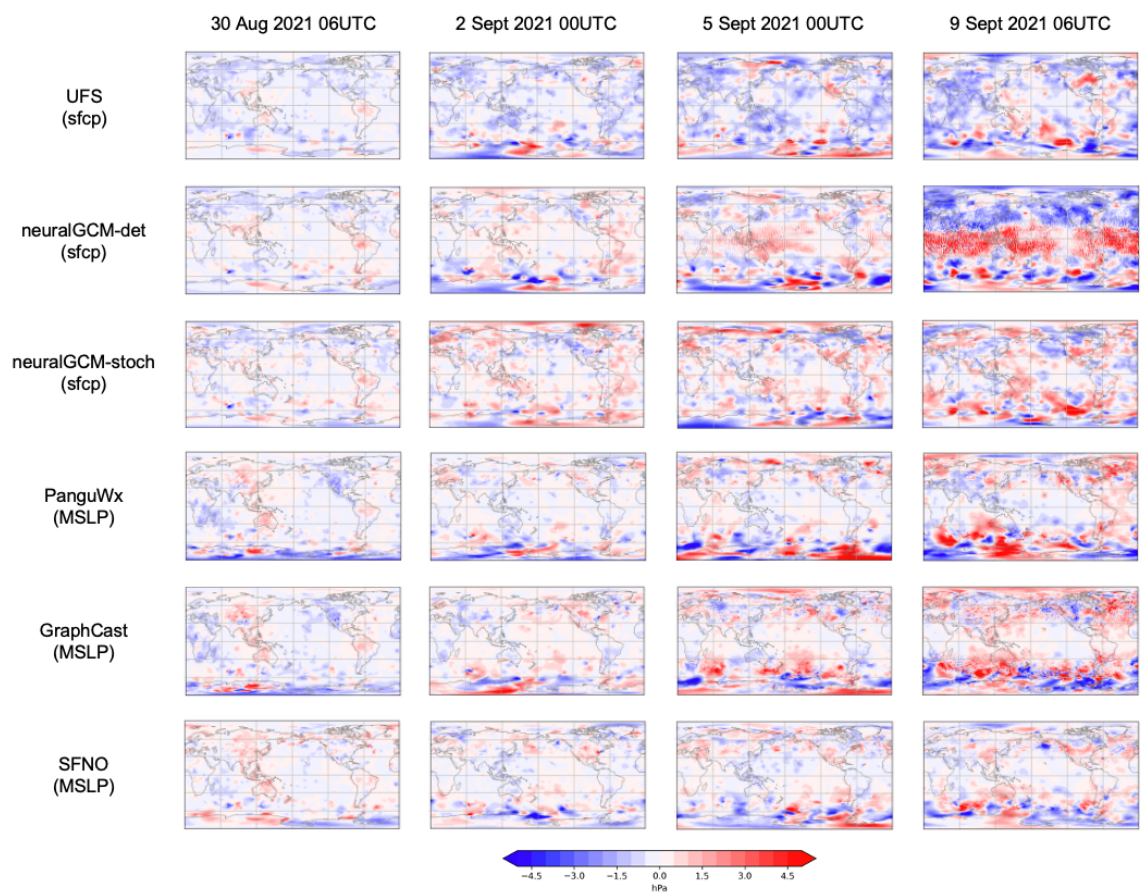

**Figure S1.** Surface pressure (sfcp) or MSLP analysis increments for four representative cycles from each experiment. Increments are expected to grow in time as the system adjusts to the sparse surface-pressure-only observing network. Note the wave-like instability in neuralGCM-det and the grid-scale noise in PanguWx, GraphCast, and (to a lesser extent) SFNO.

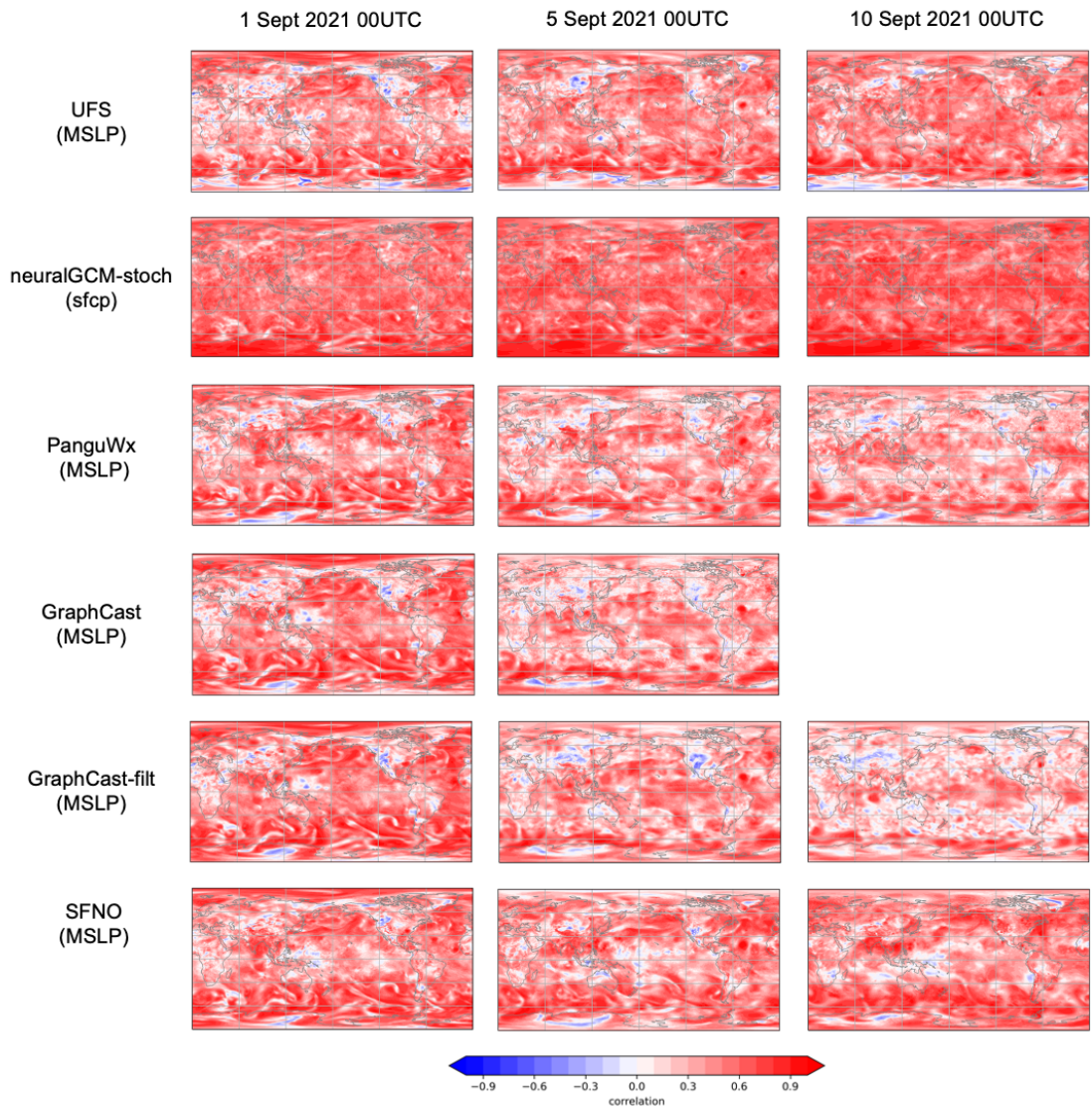

**Figure S2.** Ensemble correlations between Z500 and surface pressure (sfcp) or MSLP for each experiment, for three dates. Note that GraphCast diverged before 10 Sept.